\documentclass[final,11pt]{article}
\usepackage{multirow}
\usepackage[nottoc]{tocbibind}
\usepackage{geometry}
\usepackage{subcaption}
\usepackage[affil-it, auth-sc]{authblk}
\usepackage[english]{babel}
\usepackage[utf8]{inputenc}
\usepackage{graphicx}
\usepackage{amssymb}
\usepackage{amsthm}
\usepackage{amsmath}
\usepackage{amsfonts}
\usepackage{verbatim}
\usepackage{lineno}
\usepackage{float}
\usepackage{todonotes}
\presetkeys{todonotes}{inline, color=blue!30, size=\small}{}
\usepackage{color, soul}
\definecolor{darkred}{rgb}{0.9, 0.0, 0.0}
\definecolor{darkgreen}{rgb}{0.0, 0.5, 0.0}
\usepackage[colorlinks,bookmarks,linkcolor=darkred, citecolor=darkgreen]{hyperref}
\usepackage{eso-pic}
\usepackage[sort&compress,numbers]{natbib}
\usepackage{doi}
\usepackage{paralist,epsfig}
\usepackage{relsize}
\usepackage{nicefrac,esint}
\usepackage{cleveref}
\emergencystretch=\maxdimen
\def\slash#1{#1\!\!\!/\!\,\,}

\begin{document}

\AddToShipoutPictureFG*{
    \AtPageUpperLeft{\put(-60,-60){\makebox[\paperwidth][r]{LA-UR-22-22741}}}
}

\title{QED medium effects in (anti)neutrino-nucleus and
electron-nucleus scattering: elastic scattering on nucleons}

\author{Oleksandr Tomalak\thanks{tomalak@lanl.gov}}
\author{Ivan Vitev\thanks{ivitev@lanl.gov}}
\affil{Theoretical Division, Los Alamos National Laboratory, Los Alamos, NM 87545, USA}

\date{\today}

\maketitle

Interpretation of current and future neutrino oscillation and electron scattering experiments requires knowledge of lepton-nucleon and lepton-nucleus interactions at the percent level. We study the exchange of photons between charged particles and the nuclear medium for (anti)neutrino-, electron-, and muon-induced reactions inside a large nucleus. While quantum electrodynamics (QED)-medium contributions are formally suppressed by two powers of the electromagnetic coupling constant $\alpha$ when compared to the leading-order cross sections, low-energy modes and the nuclear size enhance the effect by orders of magnitude. They require a proper infrared regularization, which we implement as a screening of the electromagnetic interactions at atomic length scales or above. We provide approximate analytic expressions for the distortion of (anti)neutrino-nucleus and charged lepton-nucleus cross sections and evaluate the QED-medium effects for realistic values of the screening scale on the example of elastic scattering with nucleons inside the nucleus. We find new permille- to percent-level effects, which were not considered in either (anti)neutrino-nucleus or electron-nucleus scattering.

\newpage
\tableofcontents

\section{Introduction}

Electron, muon, and (anti)neutrino scattering on nucleons and nuclei provide us with fundamental information about the complex internal structure and dynamics of the target~\cite{Hofstadter:1956qs,Ernst:1960zza,Hand:1963zz,Dombey:1969wk,Akhiezer:1973xbf,Drechsel:1989ab,Boffi:1993gs,Blomqvist:1998xn,Leemann:2001dg,Arrington:2006zm,Perdrisat:2006hj,A1:2010nsl,A1:2013fsc,Bernauer:2010zga,Zhan:2011ji,Abrahamyan:2012gp,Accardi:2012qut,Qweak:2013zxf,MUSE:2013uhu,Punjabi:2015bba,Mosel:2016cwa,Aschenauer:2017jsk,Adams:2018pwt,Xiong:2019umf,Anderle:2021wcy}. This includes nucleon and nucleus electromagnetic and electroweak form factors and radii, unpolarized and polarized parton distribution functions, 3D tomography of the nucleon, and parton interactions with the nuclear environment. Electromagnetic probes are also at the heart of state-of-the-art electroweak measurements~\cite{Erler:2004in,SLACE158:2005uay,Kumar:2013yoa,Becker:2018ggl}, and utmost precision in our knowledge of (anti)neutrino-nucleus scattering is paramount for the extraction of the neutrino oscillation parameters~\cite{MINOS:2011amj,T2K:2011qtm,Hyper-KamiokandeProto-:2015xww,T2K:2019bcf,NOvA:2019cyt,DUNE:2020ypp}. These research directions call for improved description of the lepton-nucleus cross sections~\cite{Jen:2014aja,Benhar:2015wva,Ankowski:2016bji,Ankowski:2016jdd,Katori:2016yel,NuSTEC:2017hzk,Nagu:2019uco,Rocco:2020jlx,CLAS:2021neh} and, ideally, the precise first-principles microscopic theory of such interactions should be formulated and developed. The precision goal of future experiments~\cite{ESSnuSB:2013dql,Hyper-KamiokandeProto-:2015xww,DUNE:2020ypp} cannot be met without accounting for the exchange of virtual and real photons in (anti)neutrino-induced reactions~\cite{Day:2012gb,Tomalak:2021qrg,Tomalak:2022xup}. Similarly, the exchange of photons between the electrically charged leptons and nuclear medium in electron and (anti)neutrino scattering should be taken into account and has not yet been studied in the literature. In this paper, we take the first steps to address this outstanding question and show that exchange of photons with the nuclear medium can induce permille- to percent-level correction to the scattering cross sections through multiple elastic scattering on nucleons in modern and near-future electron and neutrino scattering experiments.

Understanding of particle scattering in matter is a problem of fundamental importance to many fields of science. For continuous media, a formalism to treat multiple interactions was developed by Moli\`{e}re more than seventy years ago~\cite{Moliere:1947zza,Moliere:1948zz}. In the context of quantum chromodynamics (QCD), derivation of parton scattering in large nuclei via the exchange of gluons~\cite{Gyulassy:2002yv} was motivated by efforts to understand the Cronin effect~\cite{Cronin:1974zm}, the enhancement of inclusive hadron production in proton-nucleus $p$A relative to proton-proton $pp$ collisions at transverse momenta $\mathrm{p_T}\sim$ few GeV at Fermilab fixed-target experiments. The approach was further refined to provide a boost-invariant formulation of the scattering potential generated by the matter constituents and applied to describe the broadening of back-to-back hadron correlations~\cite{Qiu:2003pm} in deuterium-gold $d\mathrm{Au}$ reactions at the Relativistic Heavy Ion Collider. Research on parton scattering in nuclear matter is still ongoing~\cite{Sadofyev:2021ohn,Ben:2022dmw,Barata:2022krd}. Here, we focus instead on electron and muon scattering off large nuclei.

The rest of our paper is organized as follows. In Section~\ref{sec:nutrino}, we derive the effects of collisional interactions in (anti)neutrino-nucleus scattering within a soft-collinear effective field theory ($\mathrm{SCET}_\mathrm{G}$) with Glauber photon exchanges between the charged leptons and the quantum electrodynamics (QED) medium at leading power. We further perform calculations without the effective field theory approximations and show phenomenological results for the cross-section corrections for several nuclei of interest to the neutrino and electron scattering communities. This formalism and results are generalized to electron-nucleus scattering in Section~\ref{sec:electron}, accounting for the QED effects prior to the hard interaction. We provide a summary and outlook in Section~\ref{sec:conclusions}. Baseline lepton-nucleon scattering cross sections upon which QED medium effects build are assembled in Appendix~\ref{app:nucleon_cross_sections}. In Appendix~\ref{app:extra_plots}, we present additional numerical examples for a wider range of beam energies relevant to experiments from T2K/HyperK to MINERvA.

\section{QED medium effects in (anti)neutrino scattering}
\label{sec:nutrino}

In this Section, we follow the effective field theory (EFT) framework that was used to derive QCD medium effects~\cite{Idilbi:2008vm,Ovanesyan:2011xy} and formulate leading QED nuclear medium contributions to the charged-current (anti)neutrino-nucleus scattering cross sections.

The charged-current (anti)neutrino-nucleus scattering provides a clear illustration of nuclear matter effects, which are induced by the exchange of photons with a nuclear medium. Only the final-state charged lepton is subject to such effects, while the four-fermion interactions (combined with QCD into the Lagrangian ${\cal{L}_\mathrm{F, QCD}}$) can be always considered as the hard scattering process. Following an elegant QCD derivation within the $\mathrm{SCET}_\mathrm{G}$ effective field theory~\cite{Idilbi:2008vm,Ovanesyan:2011xy,Rothstein:2016bsq}, we write down the Lagrangian density ${\cal{L}}_\mathrm{G}$ for QED applications:
\begin{equation}
{\cal{L}}_\mathrm{G} = {\cal{L}_\mathrm{F, QCD}} + \sum \limits_{p^\prime, p''} e^{- i \left( p^\prime - p'' \right) \cdot x} \overline{\chi}_{n^\prime, p''} i \left( n^\prime \cdot \partial + i e A^\mathrm{G} \right) \frac{\slash{\overline{n}}^\prime}{2} \chi_{n^\prime, p^\prime} + {\cal{L}^\prime_\mathrm{SCET}}. \label{eq:SCET_QED_Lagrangian}
\end{equation}
Here, ${\cal{L}^\prime_\mathrm{SCET}}$ is the Lagrangian density for power-suppressed operators and other sectors of QED  that are not used in the effective field theory calculation of this paper. The Glauber photon field is $A_\mathrm{G}$: $\left(n^\prime \cdot A_\mathrm{G} \right) = 2 \pi v \left( q \right) \delta \left( q_0 \right) e^{i q z^\prime}/e$ (in the momentum space) and is related to the static background potential $v$ arising from the unit charge at the space-time point $z^\prime$. The electric charge is $e$, and collinear spinors are $\chi$. Based on this Lagrangian, we write down the leading-order scattering amplitude $\mathrm{T}^\mathrm{LO}_\nu$ of the ``hard" charged-current neutrino scattering process $J$ at the space-time point $x_0$ as
\begin{equation}
\mathrm{T}^\mathrm{LO}_\nu = \bar{\chi}_{n^\prime,p^\prime} i J \left( p^\prime \right) e^{i p^\prime \cdot x_0}, \label{eq:LO_amplitude_neutrino}
\end{equation}
with the label momentum of the final-state charged lepton $p^\prime$ and the corresponding collinear spinor $\chi_{n^\prime, p^\prime}$. As usual, $n^\prime_\nu$ and $\bar{n}^\prime_\nu$ are conventional four-vectors that satisfy $\left(n^\prime \right)^2= \left( \bar{n}^\prime \right)^2=0$ and $n^\prime \cdot \bar{n}^\prime = 2$. The direction of the charged-lepton momentum is defined by the reference vector $n^\prime_\nu = (1,0,0,1)$. At leading power, the unpolarized scattering cross section $\sigma_\nu$ is determined by the squared matrix element as
\begin{equation}
\sigma_\nu = c_\nu \left( \bar{n}^\prime \cdot p^\prime \right) \mathrm{Tr} \left[ \frac{\slash{n}^\prime}{2} J \left( p^\prime \right) J^\dagger \left( p^\prime \right) \right],\label{eq:LO_cross_section_neutrino}
\end{equation}
where a constant $c_\nu$ depends on the kinematics of external particles.\footnote{The explicit form for the tree-level elastic lepton-nucleon scattering cross sections is presented in Appendix~\ref{app:nucleon_cross_sections}.}
\begin{figure}[tbh]
\begin{center}
\epsfig{file=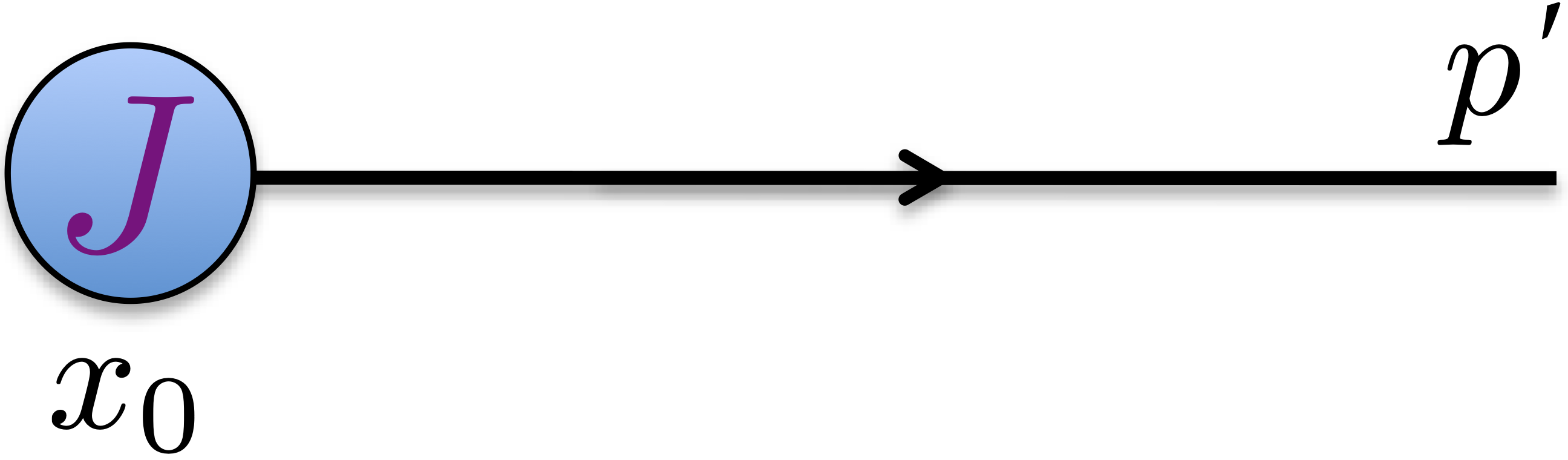, width=5.5cm}\\ \qquad\epsfig{file= 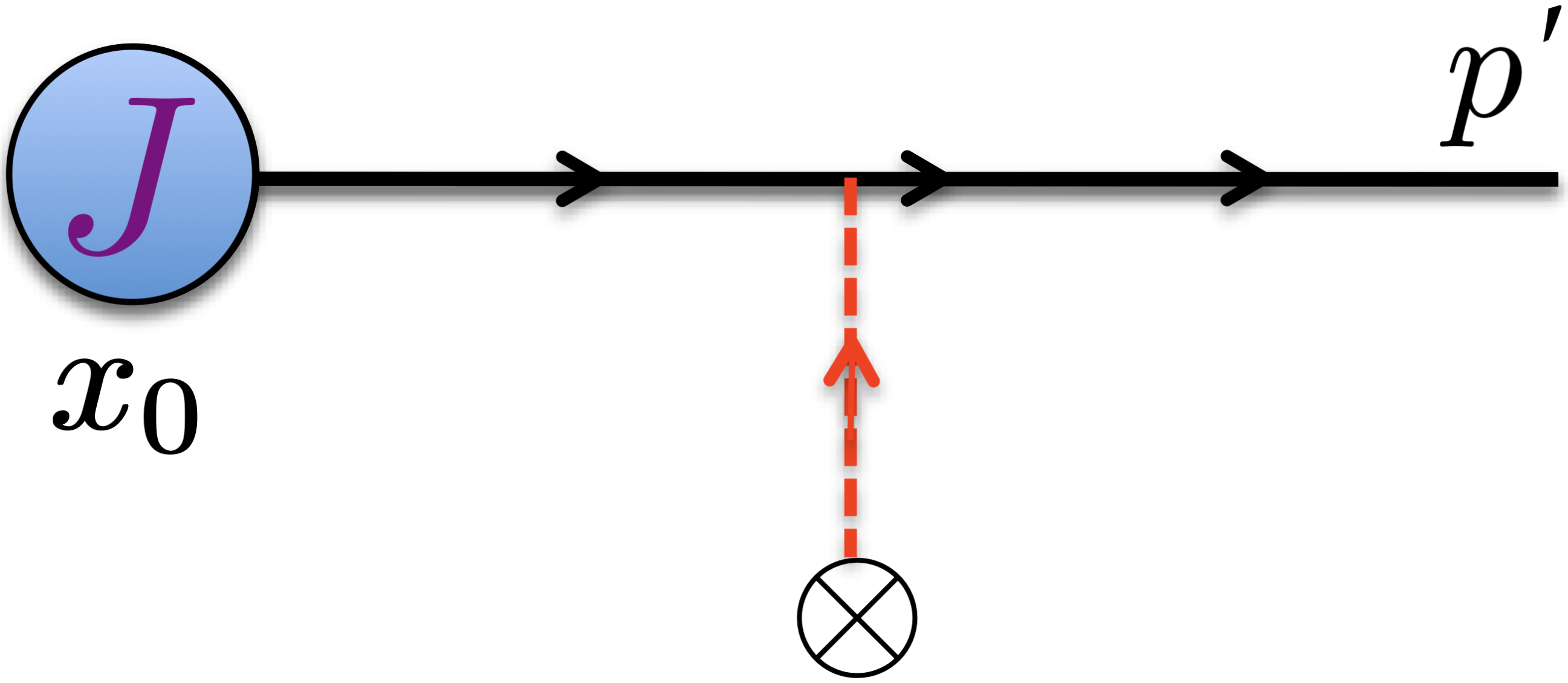, width=5.5cm} \qquad \qquad \epsfig{file= 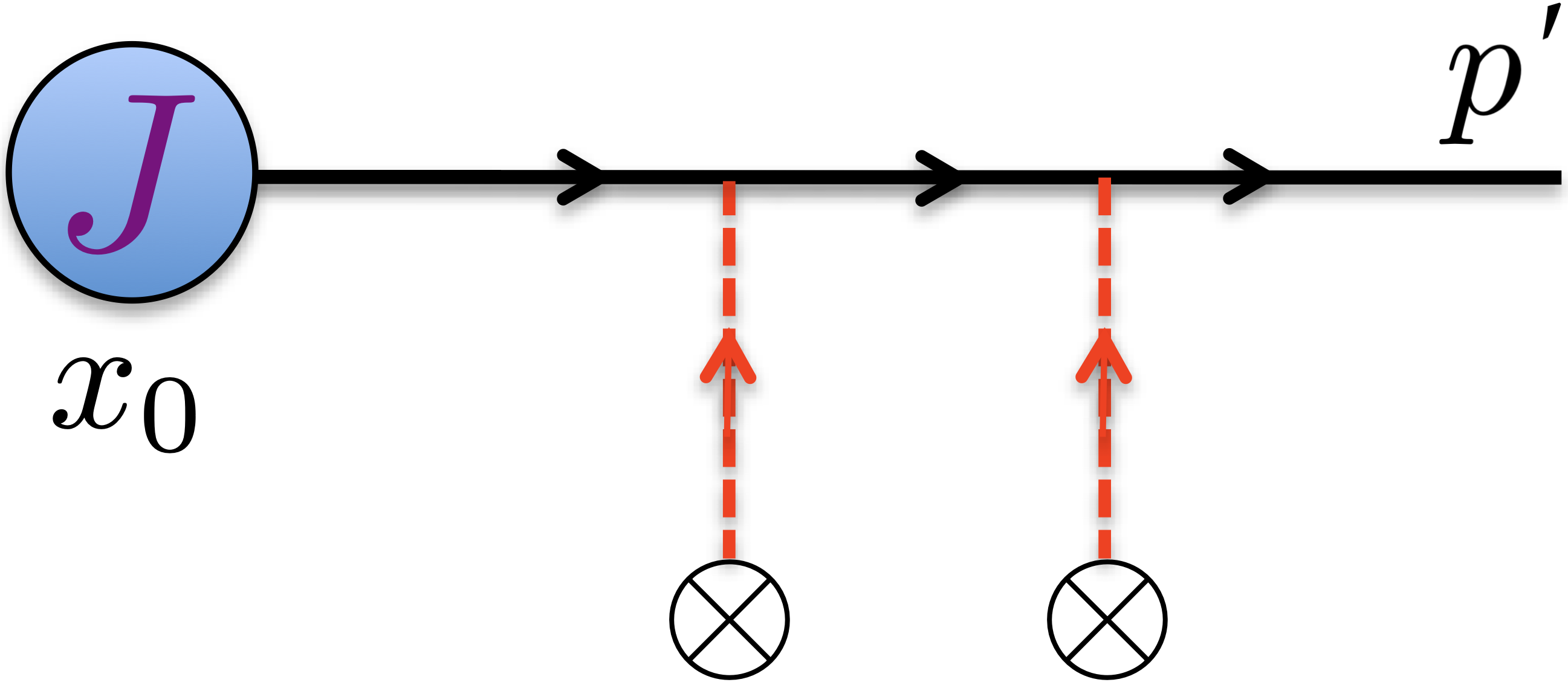, width=5.5cm}
\caption{The final-state charged lepton exchanges Glauber photons with the nuclear medium. Diagrams with one and two photons contribute to leading medium-induced corrections. \label{fig:Glauber_exchanges}}
\end{center}
\end{figure}

The electrically-charged medium serves as a source of the static electromagnetic potential, which allows us to have the same expressions for Glauber photons in an arbitrary $R_\xi$ gauge. Accounting for the exchange of one and two Glauber photons (with the dominant perpendicular component of the momentum: $q\sim \left(\lambda,\lambda,\sqrt{\lambda} \right)$ at small expansion parameter $\lambda$) between the charged lepton and the nucleus charge distribution $\rho$, as it is shown in Fig.~\ref{fig:Glauber_exchanges} (taken from \cite{Ovanesyan:2011xy}), we obtain $\mathrm{O} \left( \alpha^2 \right)$ QED medium modification of the (anti)neutrino scattering cross section $\delta \sigma_\nu$:
\begin{equation}
\delta \sigma_\nu = \int \mathrm{d} z^\prime \rho \left( z^\prime \right) \int \frac{\mathrm{d}^2 \vec{q}_{\perp}}{\left( 2 \pi \right)^2} |v \left( \vec{q}_\perp \right)|^2 \left( \sigma_\nu \left( \vec{p}^\prime - \vec{q}_\perp \right) - \sigma_\nu \left( \vec{p}^\prime \right) \right).  \label{eq:QED_medium_neutrino}
\end{equation}
In Eq.~(\ref{eq:QED_medium_neutrino}), the integration goes along the charged-lepton direction $z^\prime$, and we neglect the bending of the lepton inside the nucleus. The $\vec{q}_\perp$ integration goes over the plane that is orthogonal to the charged-lepton momentum. For typical Coulomb potential of the Glauber photon in the momentum space $v \left( \vec{q}_\perp \right) = e^2/\vec{q}^2_\perp$, the expression in Eq.~(\ref{eq:QED_medium_neutrino}) diverges in the infrared. To remove this unphysical divergence, we regularize the Coulomb potential by screening at atomic distances $R$ or above
\begin{equation}
v \left( \vec{q}_\perp \right) = \frac{e^2}{\vec{q}_\perp^2 + \zeta^2}, \qquad \qquad \zeta R \ll 1. \label{eq:Coulomb_potential}
\end{equation}

In our calculations, we assume the same Woods-Saxon spherically-symmetric distribution for neutrons and protons inside the nucleus, so that the charge distribution is expressed as
\begin{equation}
\rho \left( r \right) \sim \frac{1}{1 + e^\frac{r-R_0}{a}}, \label{eq:Woods_Saxon}
\end{equation}
with the nucleus size parameter $R_0$ and $a = 0.5~\mathrm{fm}$, and normalization to the charge of the nucleus $Z$: $Z = \int \rho \left( r \right) \mathrm{d}^3 r$ (or number of neutrons for the neutron distribution). Subsequently, we average over all nucleons, as possible scattering centers inside the nucleus. Only distributions of neutrons and protons inside the nucleus determine the dependence of medium effects on the nucleus under consideration. In this paper, we take nuclear radii from Ref.~\cite{nds_charge_radii} and collect them in Table~\ref{tab:nuclear_radii}.\footnote{Note that the root-mean-square (rms) charge radius $R_\mathrm{rms}$ is related to the parameters $R_0$ and $a$ in Eq.~(\ref{eq:Woods_Saxon}) as
\begin{equation}
R_\mathrm{rms}^2 = 12 a^2\frac{\mathrm{Li}_5 \left( - e^{\frac{R_0}{a}}\right)}{\mathrm{Li}_3 \left( - e^{\frac{R_0}{a}}\right)}. \label{eq:rms_radii}
\end{equation}}
\begin{table}[!t]
    \centering 
    \begin{tabular}{|c|c|c|c|c|c|c|c|c|c|}
    \hline
        & $^2_1 \mathrm{H}$ & $^{12}_6 \mathrm{C}$ & $^{16}_8 \mathrm{O}$ & $^{40}_{18} \mathrm{Ar}$ & $^{56}_{26} \mathrm{Fe}$ & $^{208}_{82} \mathrm{Pb}$  \\ \hline
        $R_\mathrm{rms}~(\mathrm{fm})$    & $2.1421$ & $2.4702$ & $2.6991$ & $3.4274$ & $3.7377$ & $5.5012$ \\ \hline
    \end{tabular}
    \caption{Root-mean-square charge radii for various nuclei are presented~\cite{nds_charge_radii}.}
    \label{tab:nuclear_radii}
\end{table}
Eq.~(\ref{eq:QED_medium_neutrino}) is obtained assuming the charged lepton to be collinear and massless ($m_\ell = 0$). Besides this $\mathrm{SCET_G}$ calculation, we account for all neglected power corrections by substituting into Eq.~(\ref{eq:LO_amplitude_neutrino}) exact expressions for spinors of external particles and for the charged current $J$.

In this paper, we focus on the charged-current elastic (anti)neutrino-nucleon scattering as the simplest charged-current process inside the nucleus. This process depends on two kinematic variables: the (anti)neutrino energy $E_\nu$ in the nucleon rest frame and the momentum transfer $Q^2 = - \left( p - p'\right)^2$, which we determine from the lepton kinematics solely, i.e., $p$ denotes the momentum of the initial lepton. At leading power, the shifted cross section $\sigma_\nu \left( \vec{p}^\prime - \vec{q}_\perp \right)$ in Eq.~(\ref{eq:QED_medium_neutrino}) can be obtained by the replacement of the momentum transfer $Q^2$ in the tree-level expression with
\begin{equation}
Q^2 \rightarrow Q^2 + q^2_\perp - 2 E_\nu q_\perp \sin \theta_\ell \cos \phi, \label{eq:neutrino_leading_power_replacement}
\end{equation}
where $E_\nu$ is the (anti)neutrino energy, $\theta_\ell$ is the lepton scattering angle in the nucleon rest frame, and $\phi$ is the angle of $\vec{q}_\perp$ in the orthogonal to the lepton momentum plane. The exact QED treatment includes also $\vec{q}_\perp$, $\vec{q}_{||}$ (along the lepton direction), lepton-mass, and kinematic corrections. For numerical estimates, we take the default values for the nucleon form factors from Refs.~\cite{Meyer:2016oeg} and~\cite{Borah:2020gte}. 
\begin{figure}[ht]
\centering
\includegraphics[width=0.79\textwidth]{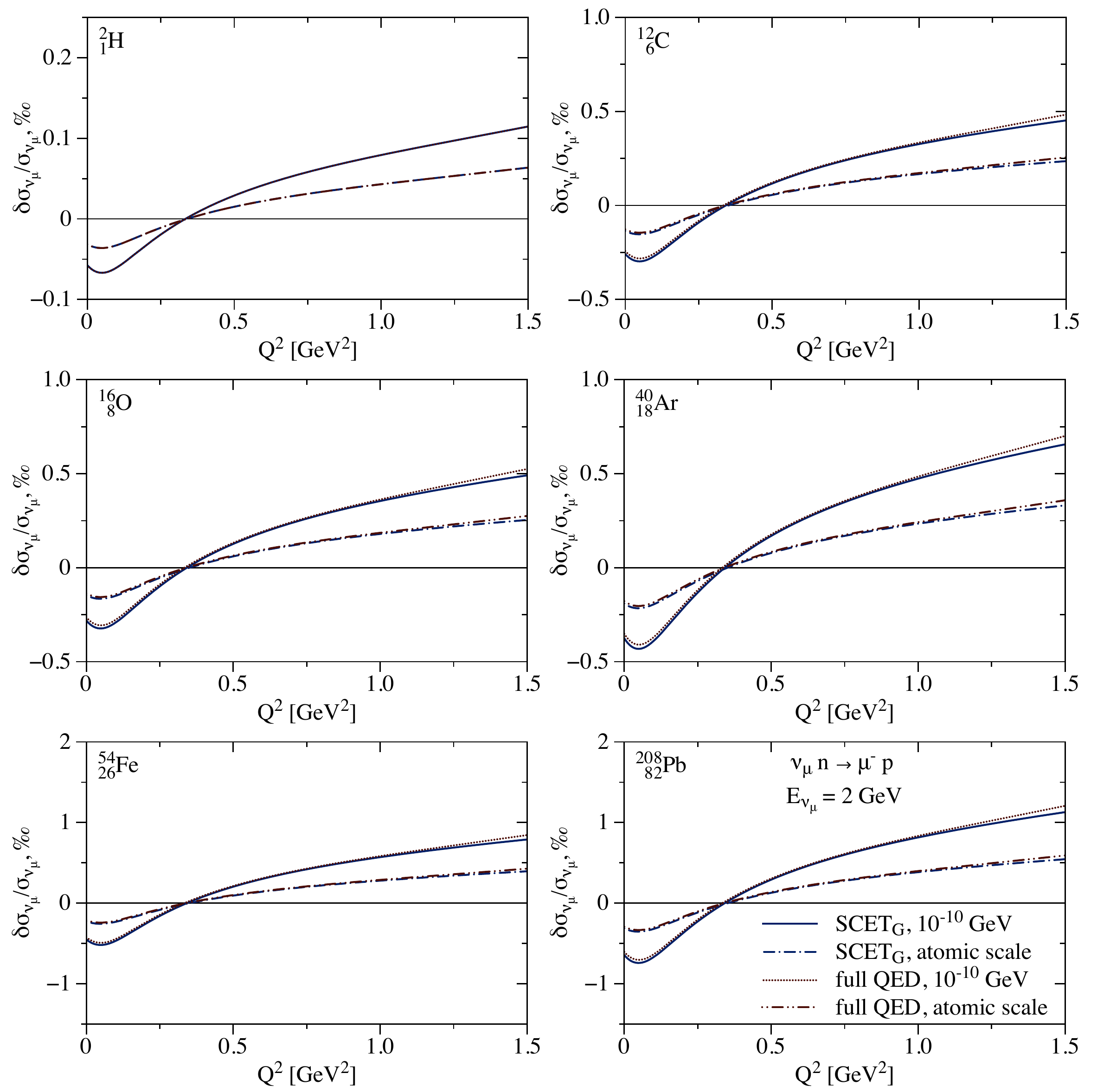}
\caption{Relative single-nucleon (inside a nucleus) $\nu_\mu n \to \mu^- p$ cross-section correction induced by QED nuclear medium effects is shown as a function of the momentum transfer $Q^2$ for the neutrino beam energy $E_{\nu_\mu} = 2~\mathrm{GeV}$. Results are presented for two values of the regularization parameter $\zeta$ in Eq.~(\ref{eq:Coulomb_potential}): $\zeta = 10^{-10}~\mathrm{GeV}$ and $\zeta = \frac{m_e \mathrm{Z}^{1/3}}{192}$. We compare the full QED result to the $\mathrm{SCET}_\mathrm{G}$ ``leading-power" approximation of Eqs.~(\ref{eq:QED_medium_neutrino}) and~(\ref{eq:neutrino_leading_power_replacement}).\label{fig:neutrino_scattering}}
\end{figure}
\begin{figure}[th]
\centering
\includegraphics[width=0.79\textwidth]{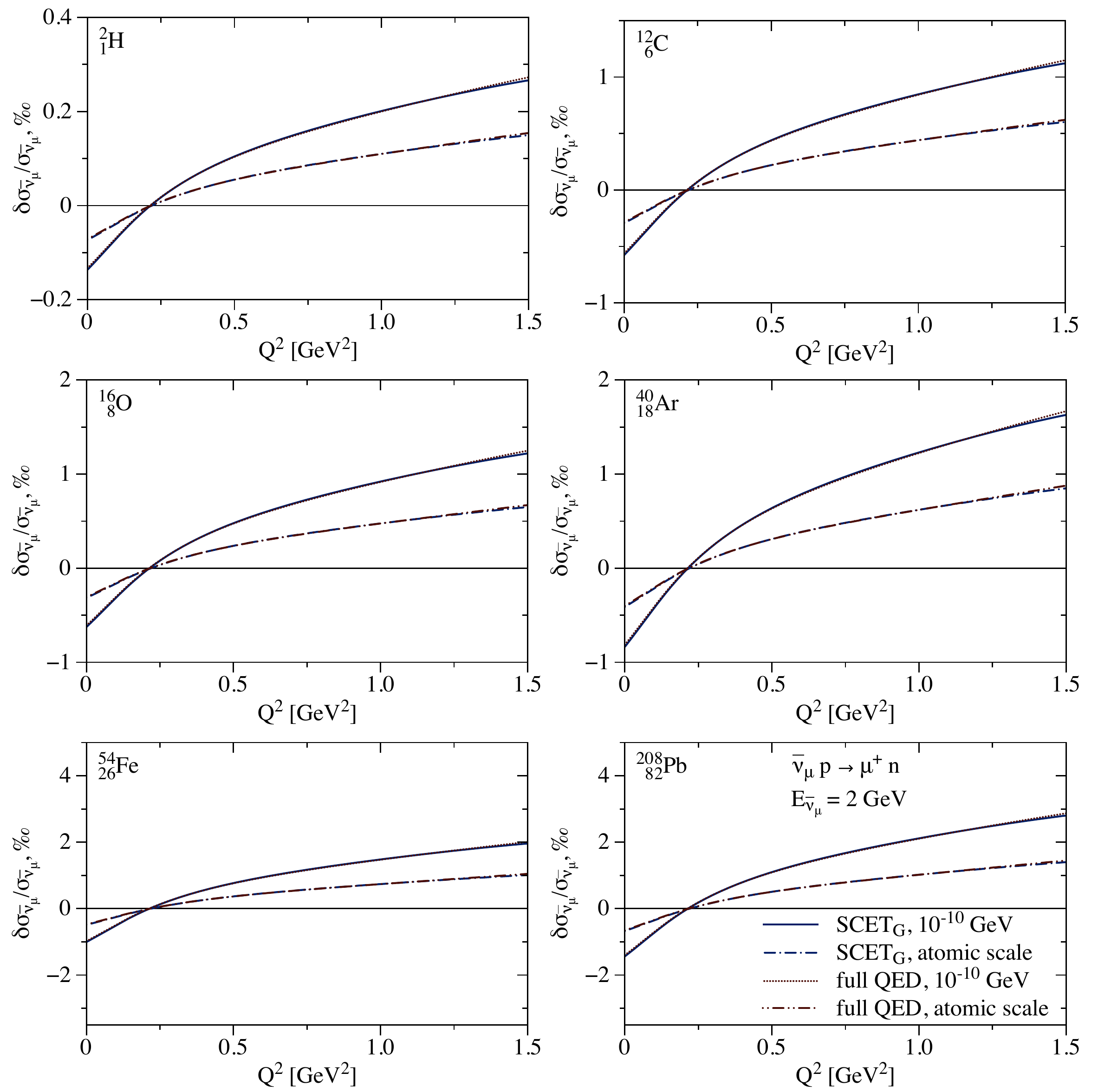}
\caption{Same as Fig.~\ref{fig:neutrino_scattering} but for the antineutrino scattering $\bar{\nu}_\mu p \to \mu^+ n$.\label{fig:antineutrino_scattering}}
\end{figure}
As illustration, we present the relative distortion of charged-current elastic cross sections per nucleon on $^2_1\mathrm{H}$, $^{12}_6\mathrm{C}$, $^{16}_8\mathrm{O}$, $^{40}_{18}\mathrm{Ar}$, $^{56}_{26}\mathrm{Fe}$, and $^{208}_{82}\mathrm{Pb}$ nuclei for the neutrino scattering in Fig.~\ref{fig:neutrino_scattering} and for the antineutrino scattering in Fig.~\ref{fig:antineutrino_scattering}. We neglect the motion and nuclear interactions of the initial and final nucleons and assume the scattering to happen on individual nucleons. We choose the (anti)neutrino beam energy $E_\nu = 2~\mathrm{GeV}$, and vary the regularization parameter between $\zeta = 10^{-10}~\mathrm{GeV}$ and the atomic screening scale $\zeta = \frac{m_e \mathrm{Z}^{1/3}}{192}$~\cite{Jackson:1998nia}. The resulting correction does not depend on the parameter $\zeta$ significantly, the variation of $\zeta$ from the atomic scale estimates the size of the uncertainty. At low momentum transfer, when the cross section is large, medium effects are negative and can reach a few permille. Going to a larger momentum transfer, the correction crosses zero and increases up to the permille-level size. Contrary to neutrino scattering, when the correction first decreases and then increases up to zero, the correction in antineutrino scattering increases monotonically at low momentum transfer due to a more rapid falloff of the tree-level cross section with the momentum transfer. The same difference in tree-level cross sections also explains why medium effects are larger for the antineutrino scattering compared to the neutrino scattering . Remarkably, the $\mathrm{SCET}_\mathrm{G}$ calculation is in very good agreement with the complete QED result, at least above the lepton-mass scale at GeV energies. Power corrections are relatively small but can be relevant for the precise first-principles cross-section calculations at high momentum transfers and (anti)neutrino energies close to the lepton production threshold. For lepton energies much larger than its mass, the size of the correction at zero momentum transfer can be easily estimated as $\frac{2 \alpha^2 \mathrm{Z}^{4/3}}{\left(M R_0 \right)^2} \left( 1 + \frac{E^\prime_\ell}{M}\right)^2 \ln \frac{\zeta}{E_\ell^\prime}$ (with the energy of the final-state charged lepton $E_\ell^\prime$ and the nucleon mass $M$), which demonstrates a logarithmic enhancement due to the separation between atomic energy scale and energies in the experiment. This simple expression estimates the order of magnitude of the correction to be at permille level or below for incoming (anti)neutrino energy $E_\nu \lesssim 3~\mathrm{GeV}$ and nuclei of interest. QED nuclear medium effects do not contain logarithms of the small lepton mass. Consequently, the corresponding corrections are (anti)neutrino flavor universal at GeV energies of the incoming beam. Phenomenological applications to lower (anti)neutrino energies and to nuclear beta decay are left to future work.

\section{QED medium effects in electron scattering}
\label{sec:electron}

In this Section, we generalize the developed formalism to the scattering of charged leptons. Both initial- and final-state leptons can interact with the medium in this case. For the hard scattering $\mathrm{H}$ at the space-time point $x_0$, we factor out spinors of initial and final states and obtain the leading-order scattering amplitude $\mathrm{T}_\ell^\mathrm{LO}$:
\begin{equation}
\mathrm{T}_\ell^\mathrm{LO} = \bar{\chi}_{n^\prime,p^\prime} \mathrm{H} \left( p^\prime, p \right) \chi_{n,p} e^{-i \left( p - p^\prime \right) \cdot x_0},\label{eq:LO_amplitude_electron}
\end{equation}
with the label momenta $p^\prime,~p$ and light-cone vectors $n^\prime,~n$ for final and initial states, respectively. For scattering of leptons, the unpolarized cross section $\sigma_\ell$ at leading power is determined by the squared matrix element as
\begin{equation}
\sigma_\ell = c_\ell \left( \bar{n}^\prime \cdot p^\prime \right) \left( \bar{n} \cdot p \right) \mathrm{Tr} \left[ \frac{\slash{n}^\prime}{2} \mathrm{H} \left( p^\prime,p \right) \frac{\slash{n}}{2} \gamma_0 \mathrm{H}^\dagger \left( p^\prime,p \right) \gamma_0 \right],\label{eq:LO_cross_section_lepton}
\end{equation}
where a constant $c_\ell$ depends on the kinematics of external particles.

Accounting for the exchange of Glauber photons with the nuclear medium, we obtain $\mathrm{O} \left( \alpha^2 \right)$ QED medium modification of the lepton scattering cross section $\delta \sigma_\ell$:
\begin{equation}
\delta \sigma_\ell = \hspace{-0.15cm} \int \hspace{-0.15cm} \frac{\mathrm{d}^2 \vec{q}_{\perp}}{\left( 2 \pi \right)^2} |v \left( \vec{q}_\perp \right)|^2 \left[ \int \mathrm{d} z^\prime \rho \left( z^\prime \right) \left( \sigma_\ell \left( \vec{p}^\prime - \vec{q}_\perp,\vec{p} \right) - \sigma_\ell \left( \vec{p}^\prime,\vec{p} \right) \right) + \int \mathrm{d} z \rho \left( z \right) \left( \sigma_\ell \left( \vec{p}^\prime, \vec{p} + \vec{q}_\perp \right) - \sigma_\ell \left( \vec{p}^\prime,\vec{p} \right) \right) \right], \label{eq:QED_medium_lepton}
\end{equation}
where the integration goes along the initial- and final-state lepton directions $z$ and $z^\prime$, respectively. We neglect the bending of leptons inside the nucleus, and specify the hard scattering process as the interaction with a larger momentum transfer. The interference of diagrams with Glauber photons attached to the initial- and final-state charged leptons vanishes exactly. Similar to (anti)neutrino scattering, we also include all power corrections by performing an exact QED calculation.

\begin{figure}[ht]
\centering
\includegraphics[width=0.79\textwidth]{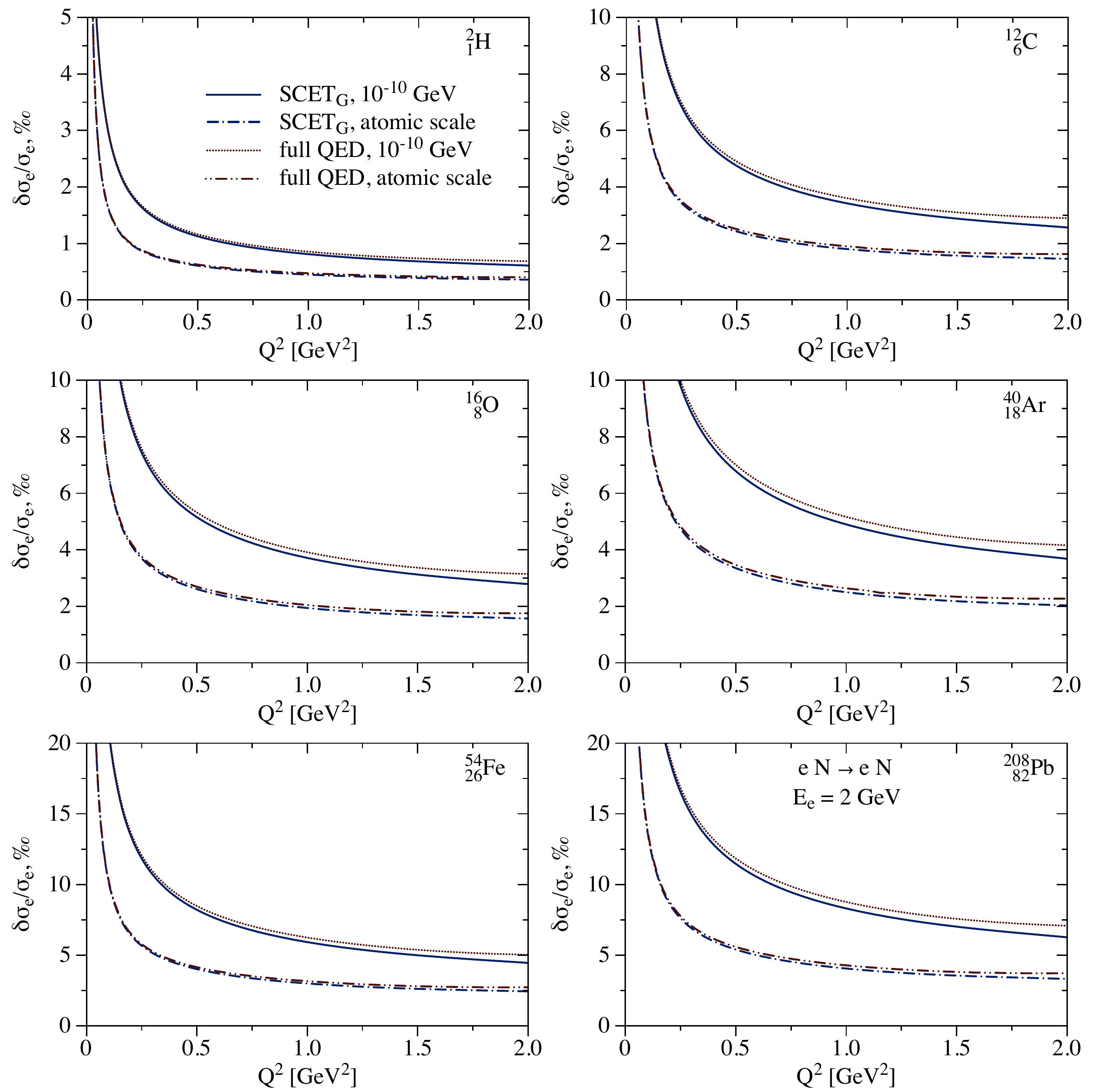}
\caption{Relative electron-nucleus cross-section correction induced by QED nuclear medium effects is shown as a function of the momentum transfer $Q^2$ for the electron beam energy $E_e = 2~\mathrm{GeV}$. Results are presented for two values of the regularization parameter $\zeta$ in Eq.~(\ref{eq:Coulomb_potential}): $\zeta = 10^{-10}~\mathrm{GeV}$ and $\zeta = \frac{m_e \mathrm{Z}^{1/3}}{192}$. We compare the full QED result to the $\mathrm{SCET}_\mathrm{G}$ ``leading-power" approximation of Eqs.~(\ref{eq:QED_medium_lepton},~\ref{eq:electron_leading_power_replacement1}), and~(\ref{eq:electron_leading_power_replacement2}). \label{fig:lepton_scattering}}
\end{figure}

For definiteness, we focus on the elastic lepton-nucleon scattering as the hard process. As all elastic $2 \to 2$ reactions, this process depends on two kinematic variables: the incoming lepton energy $E_\ell$ in the nucleon rest frame and the momentum transfer $Q^2 = - \left( p - p^\prime \right)^2$, which we determine from the lepton kinematics solely. At leading power, the shifted cross sections $\sigma_\ell \left( \vec{p}^\prime - \vec{q}_\perp, \vec{p} \right)$ and $\sigma_\ell \left( \vec{p}^\prime, \vec{p} + \vec{q}_\perp \right)$ in Eq.~(\ref{eq:QED_medium_lepton}) can be obtained by the replacement of the momentum transfer $Q^2$ as
\begin{align}
Q^2 &\to Q^2 + q^2_\perp - 2 p_\ell q_\perp \sin \theta_\ell \cos \phi,\label{eq:electron_leading_power_replacement1} \\
Q^2 &\to Q^2 + q^2_\perp - 2 p^\prime_\ell q_\perp \sin \theta_\ell \cos \phi,\label{eq:electron_leading_power_replacement2}
\end{align}
respectively. We consider the same inputs as for the (anti)neutrino-nucleon scattering and average over all nucleons that serve as scattering centers inside the nucleus. Assuming the scattering on individual nucleons and neglecting motion and nuclear interactions of the target and recoil nucleons, we present our results for the modification of elastic electron-nucleus cross sections on the nucleus in Fig.~\ref{fig:lepton_scattering}.
Contrary to (anti)neutrino interaction, the scattering amplitude of charged particles has $1/Q^2$ pole, which spoils a clear scale separation between the hard and Glauber regions. The momentum transfer in the underlying process can reach relatively small values, making the virtuality of the exchanged Glauber photon of hard-scale size. For this kinematics, we switch between the frameworks of Glauber photon exchanges from the initial- vs final-state charged lepton. We study the same nuclear targets $^2_1\mathrm{H}$, $^{12}_6\mathrm{C}$, $^{16}_8\mathrm{O}$, $^{40}_{18}\mathrm{Ar}$, $^{54}_{26}\mathrm{Fe}$, and $^{208}_{82}\mathrm{Pb}$ as for the (anti)neutrino scattering, and choose the electron beam energy $E_e = 2~\mathrm{GeV}$. Likewise, we vary the regularization parameter $\zeta$ in Eq.~(\ref{eq:Coulomb_potential}) between $10^{-10}~\mathrm{GeV}$ and the atomic scale $\zeta = \frac{m_e Z^{1/3}}{192}$~\cite{Jackson:1998nia}. The resulting correction does not depend on the parameter $\zeta$ significantly, the variation of $\zeta$ from the atomic scale indicates the size of the uncertainty. The effect grows with the atomic number of the nucleus, as it is expected. As it is mentioned above, the clear separation between hard and Glauber regions disappears going to low momentum transfers and these regions overlap at sufficiently small $Q^2$. Consequently, our correction is not well defined for $Q^2 \lesssim 0.005-0.008~\mathrm{GeV}^2$ (formally obtained when the difference of two momentum transfers in Eqs.~(\ref{eq:QED_medium_lepton}) and~(\ref{eq:electron_leading_power_replacement1}) is of order $10^{-3} Q^2$ and the slope of the correction significantly increases). In this region, the leading-order cross section and the relative nuclear medium effects diverge. Contrary to the (anti)neutrino-nucleus scattering, the dependence of this correction on the momentum transfer is monotonically decreasing, while the correction itself is positive definite. Power corrections are relatively small but might be relevant for the extraction of the nuclear spectral functions~\cite{JeffersonLabHallA:2022cit}, medium effects from this paper have to be included on top of the traditional radiative corrections and Coulomb effects, and precise first-principles cross-section calculations, especially at incoming lepton energies that are close to the threshold.

Account for the QED nuclear medium effects along the lines of the study presented in this paper can shed new light on the charge radius discrepancy between the elastic electron-proton scattering data of the recent high-precision A1@MAMI result~\cite{A1:2010nsl,A1:2013fsc} $R_p = 0.879\pm0.008~\mathrm{fm}$ and the PRad measurement~\cite{Xiong:2019umf}, with the rms charge radius of the proton $R_p = 0.838^{+0.005+0.004}_{-0.004-0.003}$, the muonic~\cite{Pohl:2010zza,Antognini:2013txn} hydrogen Lamb shift extraction $R_p = 0.84087\pm0.00039~\mathrm{fm}$, and recent measurements with ordinary hydrogen~\cite{Beyer:2017gug,Bezginov:2019mdi,Grinin1061}. However, our calculation with an approximate charge distribution works only for an estimate of the corresponding effects and points to the reduction of the scattering-data radius and toward agreement with other measurements. Applying this correction to the generated pseudo-data with a floating normalization, we obtain a reduction of the charge radius of order of the proton-radius discrepancy at hundreds $\mathrm{MeV}$ and $\mathrm{GeV}$ electron beam energies. We leave a more elaborate analysis of the electron-proton scattering data to future work.

\section{Conclusions}
\label{sec:conclusions}

In this work, we developed a method to evaluate the $\mathrm{O} \left( \alpha^2 \right)$ QED nuclear medium corrections to (anti)neutrino-nucleus and charged lepton-nucleus scattering cross sections by exploiting the $\mathrm{SCET}_\mathrm{G}$ effective field theory, and by the full QED calculation in the light-cone basis. Physically, these are the effects of multiple elastic  scattering on nucleons in the nucleus along the path of the charged-lepton propagation.  At leading power, we found that the radiation-free QED effects on (anti)neutrino-nucleus cross sections have the same form as the QCD medium corrections to quark jets. We also included the power-suppressed contributions to our numerical results. To generalize the formalism to charged-lepton scattering, we identified the necessary modifications for the exchange of photons from the incoming charged lepton and pointed out that the interference terms between the exchange of photons from the initial- and final-state charged leptons vanish, at least up to order $\mathrm{O} \left( \alpha^2 \right)$.

We presented numerical estimates on the example of the elastic scattering on nucleons embedded in various nuclei of interest to the neutrino and electron scattering communities. Our first-principles results provide thus far missing permille- to percent-level corrections to lepton-nucleus scattering cross sections. We found that QED medium effects in (anti)neutrino-nucleus scattering depend on the kinematics with suppression in the forward (small $Q^2$) direction and enhancement of the backward (large $Q^2$) scattering, while these effects in electron-nucleus scattering lead to the enhancement, which monotonically decreases from forward to backward scattering (increasing the momentum transfer).

The results presented in this paper point to sizable QED medium effects for high-Z nuclei. Our experience with strongly interacting systems indicates that medium-induced radiative corrections can be even more significant than elastic multiple scattering in nuclear matter alone~\cite{Ovanesyan:2011xy,Ovanesyan:2011kn,Fickinger:2013xwa,Kang:2016ofv} and play a key role in the modification of scattering cross sections. Calculations of the QED medium effects in pion production, deep inelastic scattering~\cite{Qiu:2004qk}, and processes with bremsstrahlung are thus an important future priority that we will pursue.

\section*{Acknowledgments}

O.T. thanks Richard Hill, Qing Chen, Kevin McFarland, and Clarence Wret for collaboration on~\cite{Tomalak:2021qrg,Tomalak:2022xup}, Richard Hill for suggestions regarding qualitative description, Kevin McFarland for suggestions regarding electron scattering section, and Weiyao Ke for useful discussions. The work is supported by the US Department of Energy through the Los Alamos National Laboratory. Los Alamos National Laboratory is operated by Triad National Security, LLC, for the National Nuclear Security Administration of the U.S. Department of Energy (Contract No. 89233218CNA000001). This research is funded by LANL’s Laboratory Directed Research and Development (LDRD/PRD) program under project number 20210968PRD4. This research was supported in part by the National Science Foundation under Grant No. NSF PHY-1748958. FeynCalc~\cite{Mertig:1990an,Shtabovenko:2016sxi}, LoopTools~\cite{Hahn:1998yk}, Mathematica~\cite{Mathematica}, and DataGraph were extremely useful in this work.

\appendix

\section{Elastic lepton-nucleon scattering cross sections at tree level \label{app:nucleon_cross_sections}}

We summarize the neutrino-neutron and antineutrino-proton charged-current elastic scattering cross sections:
\begin{align}
\nu_\ell \left( p \right) n \left( k \right) &\to \ell^- \left( p^\prime \right) p \left( k^\prime \right), \\
\bar{\nu}_\ell \left( p \right) p \left( k \right) &\to \ell^+ \left( p^\prime \right) n \left( k^\prime \right), \label{eq:neutrino_processes}
\end{align}
which get modified by the QED medium effects. At leading order in the QED coupling constant, the unpolarized differential (anti)neutrino-nucleon scattering cross section is conveniently expressed in terms of the structure-dependent $A,B$, and $C$ parameters~\cite{LlewellynSmith:1971uhs,Formaggio:2012cpf,Tomalak:2020zlv}
\begin{equation}
\frac{d\sigma_\nu}{dQ^2} (Q^2, E_\nu) = \frac{c^2_{u d}}{16\pi} \frac{M^2}{E_\nu^2} \left[ \left( \tau + r^2 \right)A(Q^2) - \nu B(Q^2) + \frac{\nu^2}{1+\tau} C(Q^2) \right] \,, \label{eq:xsection_neutrino}
\end{equation}
where $\tau = Q^2 / \left( 4 M^2 \right)$, $r = m_\ell/(2 M)$, and $\nu = E_\nu/M - \tau - r^2$. At leading order, the Wilson coefficients $c_{u d}$ of the four-fermion Lagrangian are given by $2 \sqrt{2} \mathrm{G}_\mathrm{F} V_{u d}$, where $\mathrm{G}_\mathrm{F}$ is the Fermi coupling constant and $V_{u d}$ is the Cabibbo-Kobayashi-Maskawa (CKM) matrix element. A more precise determination of the Wilson coefficients is given in~\cite{Hill:2019xqk}. Assuming the isospin symmetry, the structure-dependent factors $A,~B$, and $C$ are expressed in terms of Sachs electric, $G^V_E$, and magnetic, $G^V_M$, isovector, axial, $F_A$, and pseudoscalar, $F_P$, form factors as
\begin{align}
 A &= \tau \left( G^V_M \right)^2 - \left( G^V_E \right)^2 + (1+ \tau) F_A^2 - r^2 \left( \left( G^V_M \right)^2 + F_A^2 - 4 \tau F_P^2+ 4 F_A F_P \right) \,, \\
 B &= 4 \eta \tau F_A G^V_M \,, \\
 C &= \tau \left( G^V_M \right)^2 + \left( G^V_E \right)^2 + (1+ \tau) F_A^2 \,,\label{eq:ABC}
\end{align}
where $\eta=+1$ corresponds to neutrino scattering $\nu_\ell n \to \ell^- p$ and $\eta=-1$ corresponds to antineutrino scattering $\bar{\nu}_\ell p \to \ell^+ n$. In the limit of isospin symmetry, when the mass, $M$, for both nucleons is approximately the same, both the electric and magnetic isovector form factors are given by the difference of proton and neutron form factors, i.e., $G_{E,M}^V = G_{E,M}^p-G_{E,M}^n$. All the form factors are functions of the momentum transfer $Q^2 = - \left( p - p'\right)^2$. For electron and muon flavors, $Q^2$ increases from forward to backward directions with corresponding values $Q^2_\mathrm{-} $ to $Q^2_\mathrm{+}$, respectively,
\begin{equation}
Q^2_\pm = \frac{2 M E^2_\nu}{M+2 E_\nu} - 4M^2 \frac{M+E_\nu}{M+2 E_\nu} r^2 \pm \frac{4 M^2 E_\nu}{M+2 E_\nu} \sqrt{\left( \frac{E_\nu}{2M}- r^2 \right)^2- r^2}. \label{eq:neutirno_Q2_range}
\end{equation}

Now, we consider the elastic scattering of an electrically charged lepton on the nucleon target at rest $N$:
\begin{equation}
\ell \left( p \right) N \left( k \right) \to \ell \left( p^\prime \right) N \left( k^\prime \right),\label{eq:lepton_process}
\end{equation}
which defines the laboratory frame as $k=(M,0)$, $p=(E_\ell,\vec{p})$, $k^\prime=(M+\frac{Q^2}{2M},\vec{p}-\vec{p}^\prime)$, $p^\prime=(E_\ell- \frac{Q^2}{2M},\vec{p}^\prime)$, and the lepton scattering angle is $\theta_\mathrm{lab}$. The unpolarized differential cross section is conveniently expressed in terms of the nucleon Sachs electric and magnetic form factors $G_E$ and $G_M$ as~\cite{Preedom:1987mx,Tomalak:2018jak}
\begin{equation}
\frac{d\sigma_\ell}{dQ^2} (Q^2, E_\ell) = \frac{\pi \alpha^2}{2 M^2 \vec{p}^2} \frac{G_M^2+\frac{\varepsilon}{\tau}G_E^2}{1-\varepsilon_\mathrm{T}}, \label{eq:lepton_nucleon_cross_section}
\end{equation}
with kinematic variables
\begin{equation}
\varepsilon_\mathrm{T}=\frac{\nu_\ell^2 - \tau(1+\tau)(1+2\varepsilon_0)}{\nu_\ell^2 + \tau(1+\tau)(1-2\varepsilon_0)}\,, \qquad \quad \varepsilon=\varepsilon_\mathrm{T}+\varepsilon_0(1-\varepsilon_\mathrm{T}) \,, \qquad \quad \varepsilon_0=\frac{2m_\ell^2}{Q^2} \,, \qquad \quad \nu_\ell = \frac{E_\ell}{M} - \tau. \label{eq:lepton_nucleon_kinematics}
\end{equation}
In the elastic lepton-nucleon scattering, $Q^2$ increases from $0$ for the forward scattering to the maximally allowed value $Q^2_\mathrm{max}$:
\begin{equation}
Q^2_\mathrm{max} = \frac{4 M^2 \left(E_\ell^2 - m^2_\ell \right)}{M^2 + 2 M E_\ell + m^2_\ell}, \label{eq:lepton_nucleon_Q2_range}
\end{equation}
for the backward scattering.

\section{QED medium corrections at different beam energies \label{app:extra_plots}}

\begin{figure}[ht]
\centering
\includegraphics[width=0.79\textwidth]{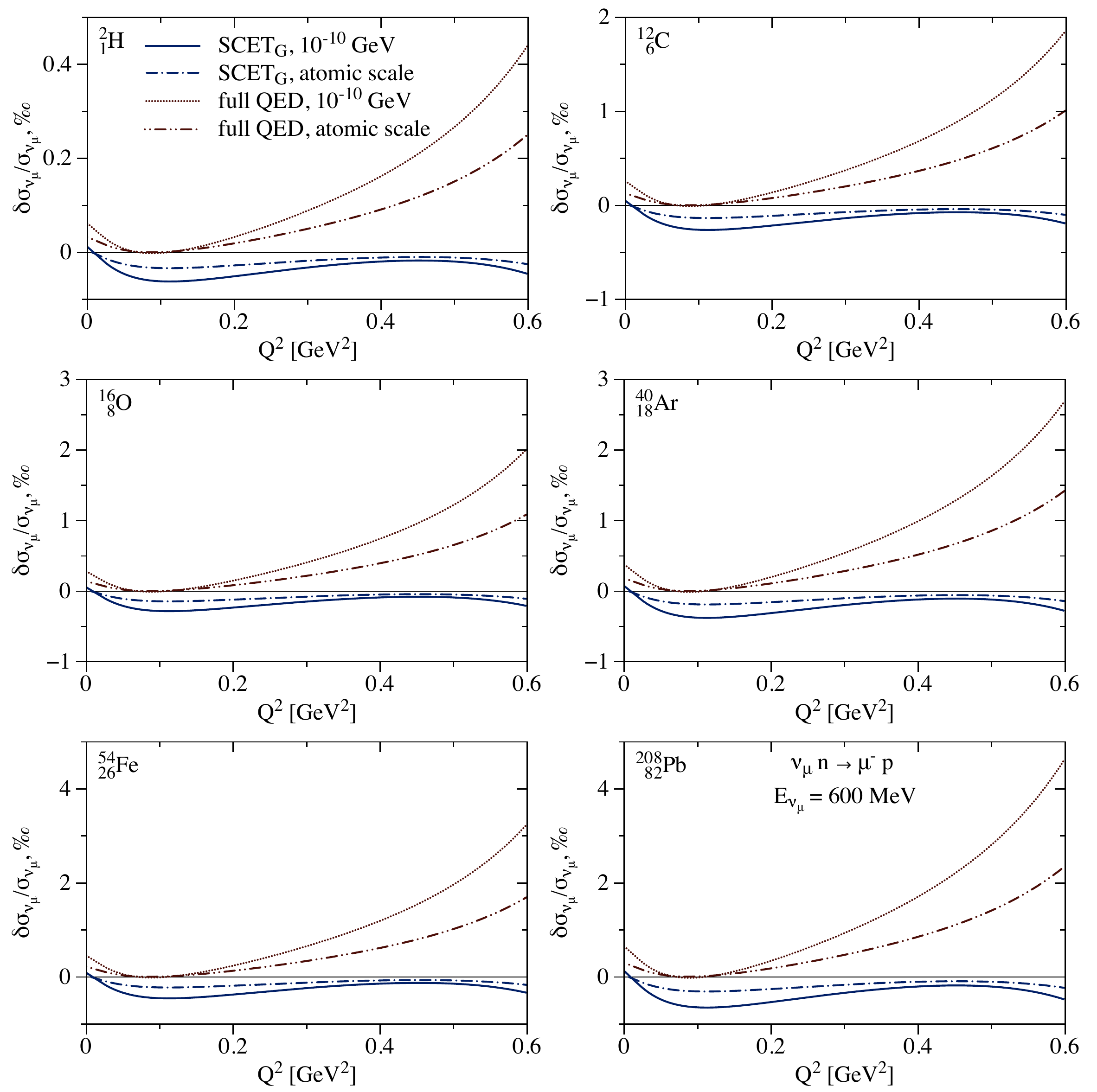}
\caption{Same as Fig.~\ref{fig:neutrino_scattering} but for the incoming neutrino energy $E_{\nu_\mu} = 600~\mathrm{MeV}$. \label{fig:neutrino_scattering06}}
\end{figure}

\begin{figure}[ht]
\centering
\includegraphics[width=0.79\textwidth]{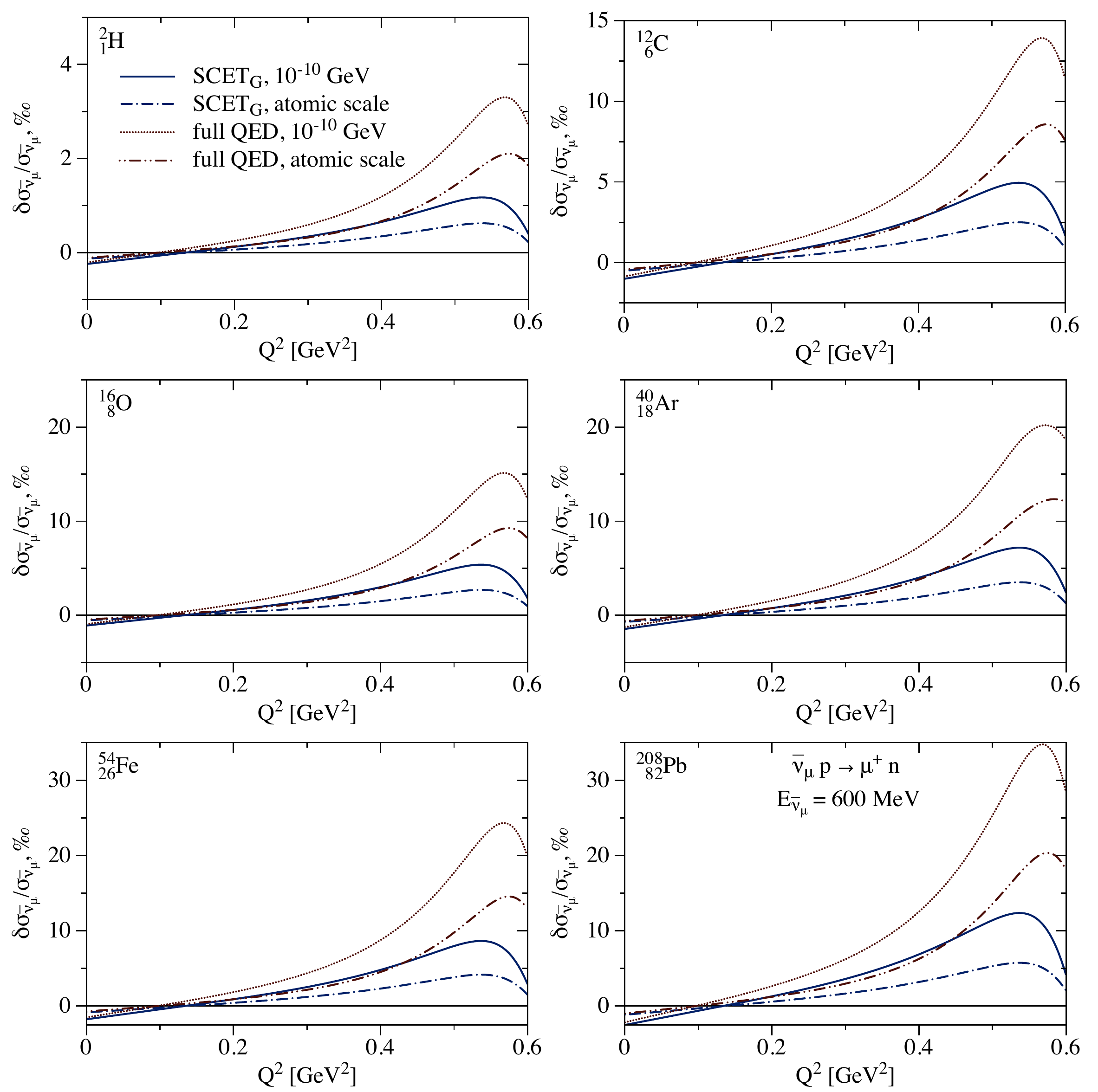}
\caption{Same as Fig.~\ref{fig:antineutrino_scattering} but for the incoming antineutrino energy $E_{\bar{\nu}_\mu} = 600~\mathrm{MeV}$. \label{fig:antineutrino_scattering06}}
\end{figure}

\begin{figure}[ht]
\centering
\includegraphics[width=0.79\textwidth]{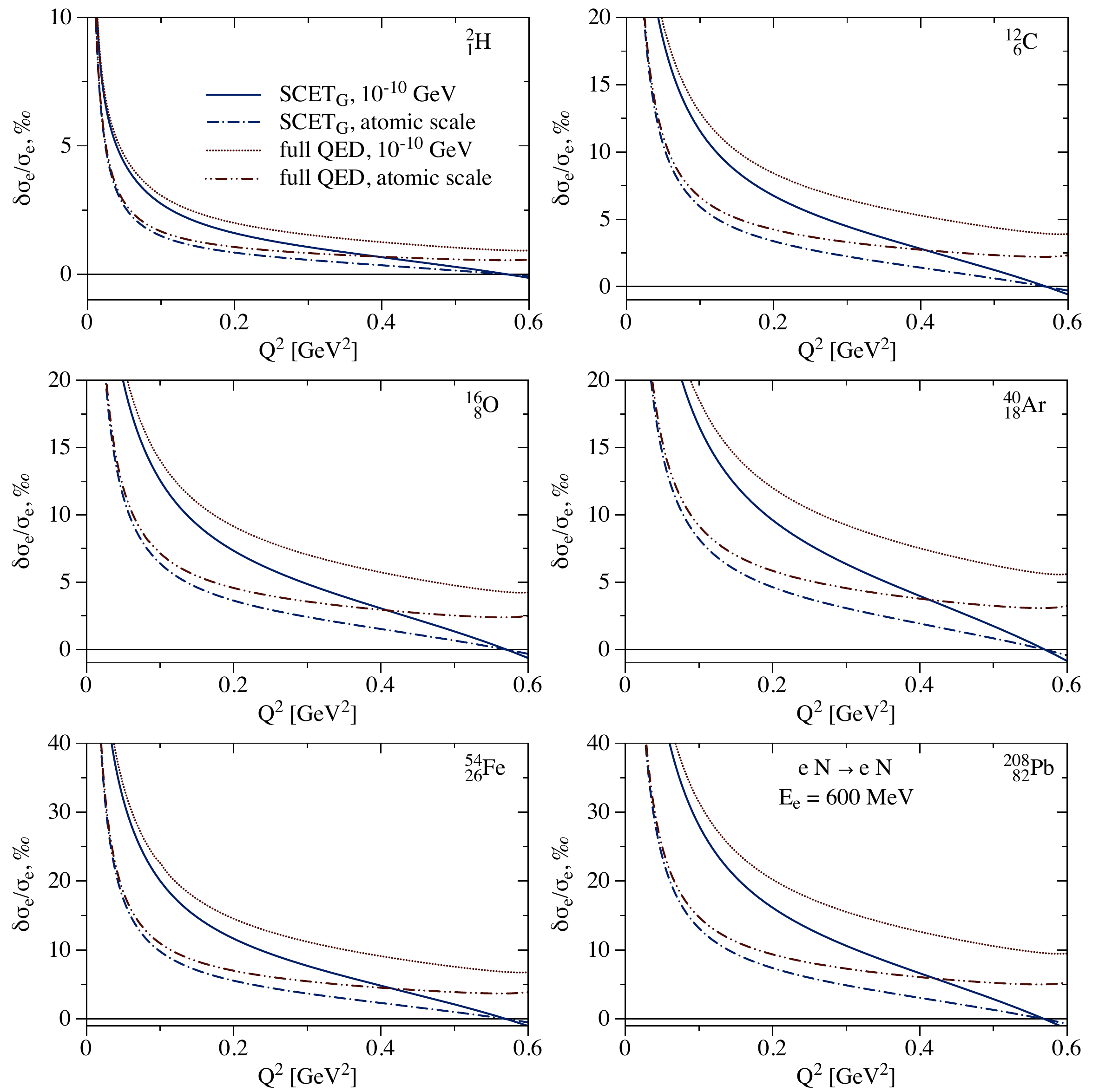}
\caption{Same as Fig.~\ref{fig:lepton_scattering} but for the incoming electron energy $E_{e} = 600~\mathrm{MeV}$. \label{fig:lepton_scattering06}}
\end{figure}

\begin{figure}[ht]
\centering
\includegraphics[width=0.79\textwidth]{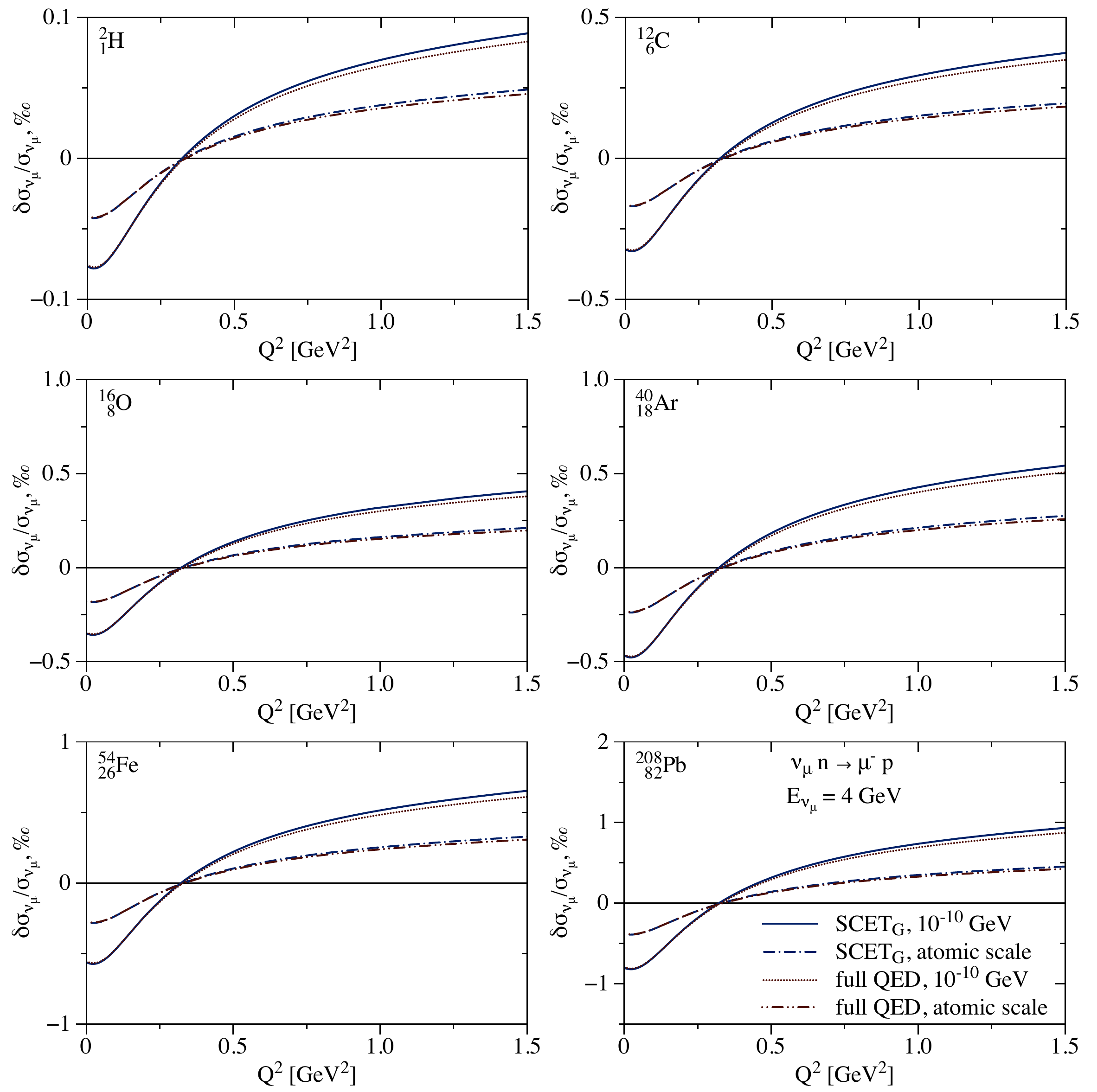}
\caption{Same as Fig.~\ref{fig:neutrino_scattering} but for the incoming neutrino energy $E_{\nu_\mu} = 4~\mathrm{GeV}$. \label{fig:neutrino_scattering4}}
\end{figure}

\begin{figure}[ht]
\centering
\includegraphics[width=0.79\textwidth]{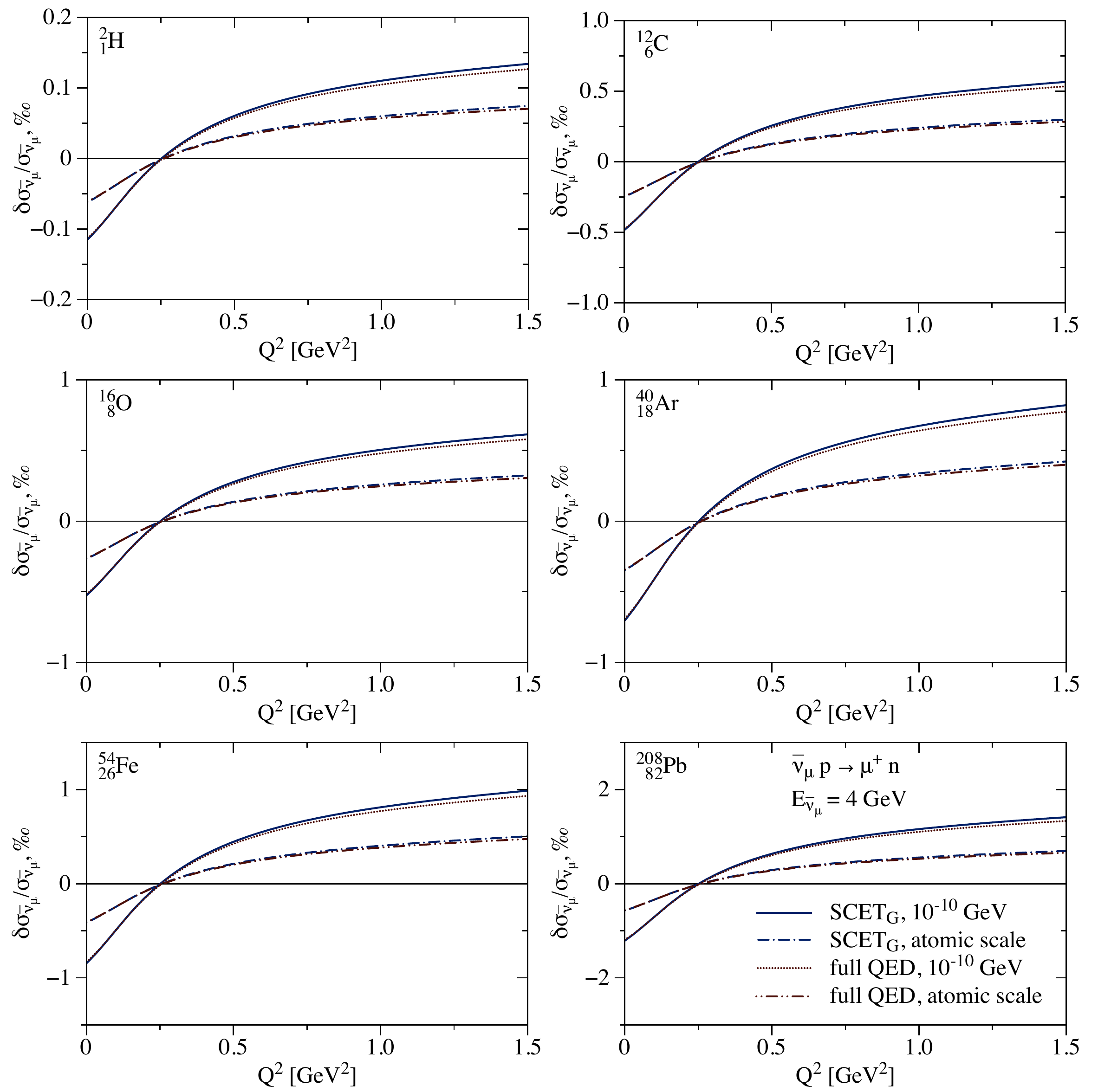}
\caption{Same as Fig.~\ref{fig:antineutrino_scattering} but for the incoming antineutrino energy $E_{\bar{\nu}_\mu} = 4~\mathrm{GeV}$. \label{fig:antineutrino_scattering4}}
\end{figure}

\begin{figure}[ht]
\centering
\includegraphics[width=0.79\textwidth]{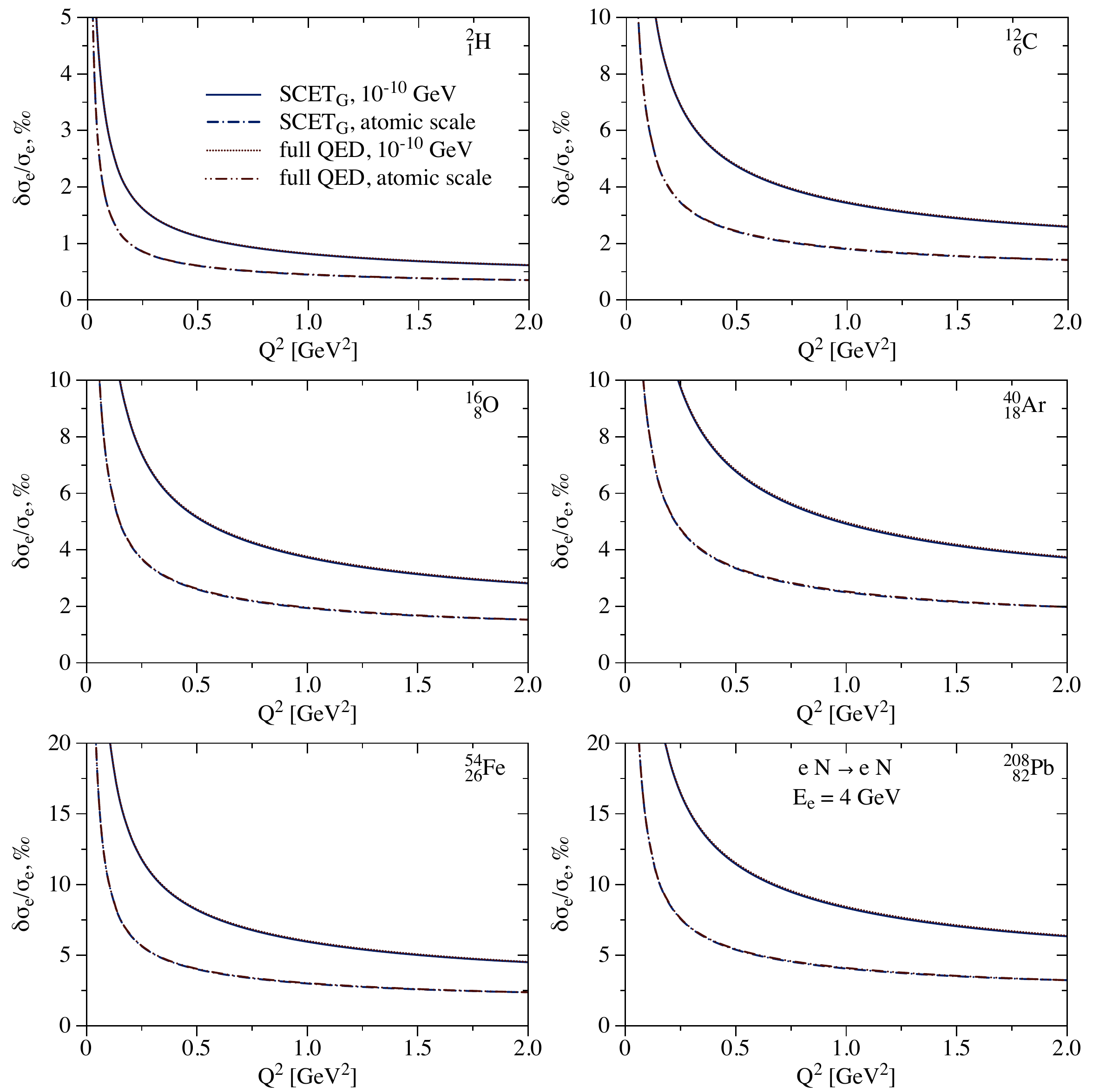}
\caption{Same as Fig.~\ref{fig:lepton_scattering} but for the incoming electron energy $E_{e} = 4~\mathrm{GeV}$. \label{fig:lepton_scattering4}}
\end{figure}

In this Appendix, we provide calculations of QED nuclear medium corrections to elastic scattering cross sections on nucleons inside the nuclei for different beam energies relevant to the experiment. Specifically, the 600~MeV choice is at the peak of T2K/HyperK~\cite{T2K:2011qtm,T2K:2019bcf,Hyper-KamiokandeProto-:2015xww} and close to the peak of short-baseline experiments at Fermilab~\cite{MicroBooNE:2015bmn,McConkey:2017dsv,Machado:2019oxb}, and we show results in Figs.~\ref{fig:neutrino_scattering06}-\ref{fig:lepton_scattering06}. We further show the 4 GeV case in Figs.~\ref{fig:neutrino_scattering4}-\ref{fig:lepton_scattering4}, which is close to the 3.5~GeV peak of the MINERvA experiment~\cite{MINERvA:2013kdn,MINERvA:2015ydy,MINERvA:2019ope,MINERvA:2017dzh,MINERvA:2018hba,MINERvA:2019gsf} and 2 times larger than our default energy. At higher energy, $\mathrm{SCET}_\mathrm{G}$ agrees with full QED calculation even better than at 2~GeV, while the lepton-mass effects and kinematic modifications become important at lower energy. Note that corrections in antineutrino scattering with $600~\mathrm{MeV}$ beam, cf. Fig.~\ref{fig:antineutrino_scattering06}, can reach $\%$ level at backward angles when the cross section itself is very small.

\bibliography{paper}{}

\end{document}